\documentclass[a4paper,11pt]{article}

\usepackage[english]{babel}
\usepackage[utf8]{inputenc}

\usepackage{geometry}
\usepackage{setspace}
\usepackage{booktabs}
\usepackage{threeparttable}
\usepackage{amsmath,amssymb,amsthm}
\usepackage{placeins}

\usepackage{graphicx}
\graphicspath{{./immages/}}
\usepackage[capposition=top]{floatrow}
\usepackage{booktabs}
\usepackage{array}
\usepackage{tabularx}
\usepackage{makecell}
\usepackage{multirow}
\usepackage{threeparttable}
\usepackage{longtable}
\usepackage{pdflscape}
\usepackage{subcaption}
\usepackage{adjustbox}
\usepackage{titlesec}
\usepackage{ragged2e}
\usepackage{changepage}
\usepackage{chngcntr}
\usepackage[colorinlistoftodos]{todonotes}
\usepackage{multirow}
\usepackage{array}
\usepackage{booktabs}
\usepackage{makecell}

\usepackage{csquotes}
\usepackage[authordate,sorting=author-chron,backend=biber]{biblatex-chicago}

\DeclareSortingScheme{author-chron}{
  \sort{\field{presort}}
  \sort{\field{sortname}\field{author}\field{editor}\field{translator}}
  \sort{\field{sortyear}\field{year}}
  \sort{\field{sortmonth}\field{month}}
  \sort{\field{sorttitle}\field{title}}
}

\DeclareFieldFormat{citehyperref}{%
  \DeclareFieldAlias{bibhyperref}{noformat}%
  \bibhyperref{#1}%
}
\DeclareCiteCommand{\citeyearwithoutparen}
  {\usebibmacro{prenote}}
  {\usebibmacro{citeindex}%
   \printtext[citehyperref]{\printfield{labelyear}}}
  {\multicitedelim}
  {\usebibmacro{postnote}}

\DeclareCiteCommand{\citeauthorwithoutparen}
  {\boolfalse{citetracker}%
   \boolfalse{pagetracker}%
   \usebibmacro{prenote}}
  {\ifciteindex{\indexnames{labelname}}{}%
   \printnames{labelname}}
  {\multicitedelim}
  {\usebibmacro{postnote}}

\usepackage[colorlinks,bookmarksopen,bookmarksnumbered]{hyperref}
\hypersetup{
  colorlinks=true,
  linkcolor=blue,
  filecolor=blue,
  urlcolor=blue,
  citecolor=blue
}

\newcolumntype{L}[1]{>{\raggedright\arraybackslash}p{#1}}
\newcolumntype{C}[1]{>{\centering\arraybackslash}p{#1}}
\newcolumntype{R}[1]{>{\raggedleft\arraybackslash}p{#1}}
\newcolumntype{P}[1]{>{\centering\arraybackslash}p{#1}}
\newcolumntype{M}[1]{>{\centering\arraybackslash}m{#1}}

\title{Identifying the effect of the US embargo on the Cuban economy: A comment on \textcite{bastos2026} \footnote{I thank Eric Freeman, Dorothy Kronick, Alex Main, Guillaume Long, and David Rosnick  for valuable comments and suggestions, and Giancarlo Bravo, Luisa García and Angel Piñango for excellent research assistance. All errors remain my own.}}
\author{\Large Francisco Rodríguez\footnote{Center for Economic and Policy Research and Josef Korbel School of Global and Public Affairs, University of Denver. E-mail: francisco.rodriguez4@du.edu}}

\begin{document}
\maketitle
\onehalfspacing

\begin{abstract}
\textcite{bastos2026} argue that the US embargo explains less than one-tenth of the difference in per capita income between Cuba and a counterfactual scenario in which the country did not follow socialist economic policies. We show that their results are driven by the use of an elasticity of income to trade openness that is neither representative nor a reasonable upper bound of the values found in the literature and by their choice to attribute the effect of the interaction between the embargo and other determinants of growth solely to those other determinants. We show that, once these problems are corrected, the embargo can account for a substantial fraction, and in some cases all, of Cuba’s post-1959 economic underperformance.  \end{abstract}

\newpage

\section{Introduction}

\textcite{bastos2026} (henceforth BGB) study the effect on Cuban living standards of events affecting the country's economy on or shortly after 1959. Using a synthetic control approach, they compare the evolution of Cuban GDP per capita with that of a counterfactual Cuba and find a large divergence in living standards from 1959 on. In their baseline estimate, Cuba's per capita income in 1989 is 56\% lower than the counterfactual. 

BGB also claim to distinguish between the effects of three different  factors that could have had a causal incidence on Cuban living standards after 1958: the socialist policies of the Cuban Revolution, the US economic embargo, and Soviet subsidies.  The authors calculate an upper bound of the effects of the US economic embargo based on estimates that, in their words, are ``deliberately less realistic", ``generous to claims about the embargo's damages", and ``constructed to overstate its impact" [BGB, 5].  Based on these calculations, they conclude that the embargo explained at most 8\% of the difference between the actual and counterfactual GDP per capita.

We find that BGB's calculations, far from providing an upper bound to the embargo's effects, are on the lower end of the range of estimates that can be derived by using elasticities of income to openness estimated in some of the most well-known contributions to the literature.  While BGB cite three references supporting their elasticity choices, we find that only one of them supports the value that they use.  We also show that BGB adopt a decomposition which attributes any growth effect of the interaction between the embargo and other growth determinants purely to those other determinants.  This decision systematically biases down their estimates of the embargo effects.

This paper provides calculations of the growth effects attributable to the embargo from BGB's synthetic control exercise that address these two problematic choices.  We show that once one uses elasticity estimates derived from some of the contributions ignored by BGB and adopt an appropriate additively separable decomposition, the embargo can explain a sizable share, and in some cases the totality, of the country's economic underperformance. 

The rest of this note is organized as follows. Section 2 explains BGB's methodology for assessing the contribution of the embargo to Cuban economic growth in the post-revolutionary period and the role played by the elasticity of income to trade and separability assumption in that estimate.  Section 3 provides an overview of the most prominent contributions to the literature on trade shares and economic growth and shows the sensitivity of the BGB results to the choice of parameter and specification choice.  Section 4 provides concluding comments.
\section{How BGB estimate the embargo's contribution to Cuban GDP}
At the center of BGB's approach to estimating the embargo's effect on Cuban living standards is a comparison of the Cuban economy's realized path of GDP per capita with that of a counterfactual Cuba built using synthetic control methods.  The authors recognize that the synthetic control exercise in itself is unable to disentangle the effects of three different factors that may have affected living standards in the post-revolutionary period: socialist economic policies, the US economic embargo, and trade subsidies from the Soviet Union.  They use estimates from \textcite{devereux2021} to deduct the amount of Soviet subsidies from Cuban GDP as well as to correct for possible over-reporting of economic growth in official data.  Assuming that both of these adjustments are correct, this leaves the task of disentangling the effects of socialist policies from the US economic embargo.

BGB's approach to this task is to bound the possible effect of the embargo by multiplying the drop in Cuban trade openness that followed the embargo by an estimate of the effect of trade openness on economic growth derived from the literature.  Formally, BGB calculate the change in income per capita between times $T_0$ and $T_1$ as:

\begin{equation}
\label{eq1}
    \Delta y=y_{0}((1+\varepsilon\Delta\lambda)^{T_1-T_0}-1)
\end{equation}

where $y$ is income per capita, $\lambda$ is trade openness defined as exports plus imports over GDP,  $\varepsilon$ is the semi-elasticity of income growth to trade openness, $\Delta x=x_{T_1}-x_{T_0}$ for any variable $x$, and $T_0$ and $T_1$ denote respectively the initial and terminal time periods.

In order to estimate a range of effects for the embargo using equation (\ref{eq1}), BGB use three estimates of $\Delta\lambda$: the decline in trade with the rest of the world relative to the synthetic control, the decline in trade with the United States relative to the pre-embargo level, and the sum of the decline in trade with the US relative to the pre-embargo level and the difference between synthetic control exports and initial exports.\footnote{Note that the way in which BGB define some of these effects is less than straightforward from their text. For example, they write that their third counterfactual scenario "is derived by replicating our analysis [used to construct the second scenario] using only exports to the United States rather than total trade" [p. 28]. This would seem to suggest that their measure of trade shock for this scenario is given by the decline in Cuban exports to the US as a percent of GDP. This would imply a shock of lower magnitude than that of their second scenario, since the decline in exports to the US is smaller than the decline of the sum of exports and imports to the US.  Such a straightforward interpretation, however, would be contradictory with their finding of larger growth effects in their third than in their second scenario, as well as their claim that the third scenario is their "most exaggerated" estimate.  As a replication package for their paper is not currently available, we derive our understanding of their methods from being able to exactly replicate their reported results.  Details of our replication are provided in the appendix.}

\subsection{Choice of Elasticity}

All of BGB's calculations use the same value for the semi-elasticity of growth to openness, $\varepsilon=0.018$.  This value implies that a one-percentage point increase in the trade ratio (e.g., an increase from 55\% to 56\%) leads to an increase of 0.018 percentage points in the growth rate of per capita income.  They cite three sources for this value: \textcite{Yanikkaya2003}, \textcite{raghutla2020} and \textcite{heimberger2022}.  However, two of these sources do not support their claim. 

 BGB state that \textcite{raghutla2020} finds that ``a 10 point increase in trade openness is associated with a 0.186\% increase in economic growth" and that the coefficient is similar to \textcite{Yanikkaya2003}'s estimate of $\varepsilon=0.018$.  This is not what Raghutla finds. \textcite[6]{raghutla2020} estimates an elasticity of 0.186 of the level of income to the trade share, indicating that a one percent increase in the trade share leads to an increase of 0.186 percent in the level of income. 
 
The Yanikkaya and Raghutla coefficients are not comparable for at least two reasons. The first one is that Yanikkaya estimates a semi-elasticity derived from a semi-logarithmic functional form while Raghutla estimates an elasticity derived from a fully logarithmic specification. The second one is that Yanikkaya estimates a short-term effect derived from a specification which explicitly models transitional dynamics, while Raghutla estimates a long-term elasticity derived from a cointegration relationship. 

Even if we ignore these problems, it is unclear how BGB arrive at the conclusion that the coefficients  are ``highly similar".  Raghutla's elasticity estimate of 0.186 is more than ten times Yanikkaya's semi-elasticity estimate of 0.018. Evaluated at $\lambda=.55$, Cuba's pre-revolution openness, a one-percentage point increase in openness in Raghutla's model would lead to a 0.34 percent increase in GDP, more than 19 times the effect that would be derived from the Yanikkaya specification. 

By controlling for a lagged dependent variable, specifications such as that of \textcite{Yanikkaya2003} capture the effect of a shock to openness on income at time $t$ holding fixed the initial level of income. This effect is transmitted gradually to future increases in income through the effect of GDP at time $t$ on GDP at time $t+1,t+2,$ and so on, with the economy eventually converging to a new steady-state level of income.  In contrast, in specifications such as that of \textcite{raghutla2020}, the coefficients capture the permanent effect of openness on income.\footnote{\textcite{raghutla2020} applies cointegration tests and estimates the relationship by fully modified OLS, explicitly referring to his estimates as "long-run elasticities" [p. 6].}  It is thus not surprising that \textcite{raghutla2020}'s coefficient estimate is so much larger than that of \textcite{Yanikkaya2003}.  

More formally, \textcite{Yanikkaya2003}'s estimate is a variant of the empirical growth  specification pioneered by  \textcite{barro1991}, namely:
\begin{equation}
\label{barro}
    \frac{\Delta y}{y_0}=\alpha_0+\alpha_1\ln y_0+\alpha_2 \lambda + \beta X_1
\end{equation}

whereas \textcite{raghutla2020} estimates a levels specification in logarithms:
\begin{equation}
\label{levels}
    \ln y_t = \gamma_0 + \gamma_1\ln \lambda_t + \theta X_2
\end{equation}

where $X_1$ and $X_2$ denote sets of additional controls. Using the logarithmic approximation to growth rates we can rewrite (\ref{barro}) as:

\begin{equation}
\label{barro1}
    \ln{y_t}=\alpha_0+(\alpha_1+1)\ln y_0+\alpha_2 \lambda + \beta X_1
\end{equation}

There are thus two key differences between the growth specification used by \textcite{Yanikkaya2003} and the levels specification of \textcite{raghutla2020} (aside from the effect of different sets of additional controls). The first one is that equation (\ref{levels}) uses the logarithm of openness as the explanatory variable whereas equation (\ref{barro1}) uses the level of openness.  The second one is that equation (\ref{barro1}) controls for the log of initial GDP whereas equation (\ref{levels}) does not.  In other words, Yanikkaya estimates the effect of openness on income at the end of a given time interval, holding the initial level of income fixed, while Raghutla estimates the permanent effect of openness on the level of income.  We return to this distinction at the end of this section.

The second authority used by BGB to validate their use of $\varepsilon=0.018$  is \textcite{heimberger2022}, which they argue supports their claim that the \textcite{Yanikkaya2003} estimate is inherently conservative. In their words: 
\begin{quote}
``we used this estimate for safety reasons. Other literature, including meta-analyses
such as that of Heimberger (2022), suggest smaller estimates. Using a larger estimate is meant to be more generous towards claims that the embargo had an effect so large that it is misattributed to the Revolution." [BGB, p. 28, fn. 40]
\end{quote}

\textcite{heimberger2022}, however, provides no support for this claim.  While Heimberger's paper does provide a meta-analysis of the literature on the effect of globalization on growth, its analysis is completely based on a comparison of partial correlation coefficients calculated using only reported t-statistics and degrees of freedom.  \textcite[1695]{heimberger2022} explicitly defends his choice to center on a measure of statistical association by arguing that, in contrast to absolute effect sizes, it ``can be meaningfully compared across papers (because it is a unitless measure bounded between -1 and 1), and ...can be calculated for a much larger set of estimates than other effect-size measures."  His paper is therefore uninformative on the issue of whether $\varepsilon=0.018$ found by \textcite{Yanikkaya2003} is higher or lower than that of other studies in the literature. 

An additional problem with BGB's implementation of equation (\ref{eq1}) is that it is obviously sensitive to the choice of terminal time period $T_1$. BGB choose $T_1=1972$, 10 years after the implementation of the full embargo, but make no attempt to justify this horizon.  To the extent that the objective of their investigation is to understand the sources for the differences in Cuban living standards with respect to a counterfactual in 1972, that could be a defensible choice. However, if the aim is to understand the long-run causes of Cuba's economic underperformance, the choice 
appears arbitrary.  

As we discuss in Section \ref{levels}, the bulk of recent empirical estimates of the elasticity of income with respect to openness uses a levels specification. In that case, estimated coefficients—of which \textcite{raghutla2020} is an example—capture long-term, permanent effects. Such coefficients appear more adequate than short-term elasticities for the purposes of capturing the long-run effects of the embargo on Cuba's economic development. 

Long-term effects can also be derived from Barro-style growth specifications with a lagged dependent variable like that used by Yanikkaya.  Given $\alpha_1<0$, equation (\ref{barro1}) defines a stable difference equation with a single positive steady state level of income, $y_{ss}$. We can derive the semi-elasticity of $y_{ss}$ to $\lambda$ by setting the growth rate to 0 and differentiating, which gives:
\begin{equation}
\label{ss}
    \frac{1}{y_{ss}}\frac{dy_{ss}}{d\lambda}=-\frac{\alpha_2}{\alpha_1}.
\end{equation}

\subsection{Separability}
\label{secdec}
BGB implicitly assume that the effects of the embargo and socialist policies compound geometrically yet are additively separable in levels.  Those two assumptions are generally incompatible. To see why, assume that income is determined according to the equation:
\begin{equation}
\label{deflny}
    \ln y = \alpha_0 + \alpha_1\lambda+\alpha_2 \eta
\end{equation}
where $\lambda$, as before, denotes trade openness and $\eta$ is denotes other factors influencing growth, such as policies.  For the purposes of this discussion, we will assume that BGB's synthetic control exercise is successful at controlling for all other factors different from the embargo and socialist policies. In that case the difference between the logs of  synthetic counterfactual and  realized income  will be given by:
\begin{equation}
\label{addsep}
    \ln y^{NE,NS}- \ln y^{E,S} = \alpha_1 (\lambda^{NE}-\lambda^{E})+\alpha_2(\eta^{NS}-\eta^S),
\end{equation}
allowing us to calculate the effect of the embargo relative to the total underperformance of the economy as:
\begin{equation}
\label{share}
    \theta=\frac{c^{NE}}{c^{NE}+c^{NS}}
\end{equation}
where
\begin{equation}
    c^{NE}=\alpha_1(\lambda^{NE}-\lambda^{E});c^{NS}=\alpha_2(\eta^{NS}-\eta^S).
\end{equation}

However, BGB calculate a different expression, which is:
\begin{equation}
\label{thetaBGB1}
\theta_{BGB}=\frac{y^{NE,S}-y^{E,S}}{y^{NE,NS}-y^{E,S}}    
\end{equation}
This expression can be rewritten as:

\begin{equation}
\label{thetabgb}
 \theta_{BGB}=   \frac{g^{NE}}{g^{NS}+g^{NE}+g^{NS}g^{NE}}
\end{equation}
where 
\begin{equation}
\label{defs}
    g^{NE}=e^{\alpha_1(\lambda^{NE}-\lambda^E)}-1;g^{NS}=e^{\alpha_2(\eta^{NS}-\eta^S)}-1
\end{equation}

In contrast to its counterpart in logarithms, $\theta_{BGB}$ does not  provide us with an additively separable decomposition (i.e., one where the total effect can be written as a sum of the embargo and policy effects). The denominator of the right-hand side of equation (\ref{thetabgb}) includes the term $g^{NS}g^{NE}$ which represents the interaction between lifting the embargo and ending socialist policies and cannot cleanly be attributed to either factor. BGB completely apportion this term to non-embargo effects, a decision which biases down the estimated effect of the embargo when $g^{NS}$ and $g^{NE}$ are both positive. 

In other words, going from the current situation of Cuba to that represented by the synthetic control would require two changes in conditions: lifting the embargo and adopting a market system.  Each of them will raise growth if done individually, while there will be an additional effect on growth if they are done together. BGB assume that this additional growth effect should be attributed only to market reforms and not to the embargo.

A cleaner and less arbitrary solution is to use (\ref{share}) to decompose the effects of the embargo and socialist policies in the space in which they are naturally additively decomposable. We adopt this choice in the next section and show how it increases the estimated embargo effects.

It is worth asking under what conditions BGB's decomposition of the embargo and socialist policies effects would be additively separable.  Let
\begin{equation}
\label{lineary}
    y = \alpha_0 + \alpha_1\lambda+\alpha_2 \eta.
\end{equation}

Then (\ref{thetaBGB1}) becomes:
\begin{equation}
    \theta_{BGB}=\frac{\alpha_1(\lambda^{NE}-\lambda^E)}{\alpha_1(\lambda^{NE}-\lambda^E)+\alpha_2(\eta^{NS}-\eta^S)}.
\end{equation}
Additive separability would therefore arise from a specification in which the level of income, and not its logarithm, is linear in trade shares and socialist policies.  This would imply that the effect of a change in trade shares (or in policies) would be independent of the initial level of income.  In that case, the effect on per capita incomes of a trade embargo would be to decrease per capita incomes by, say, \$3,000 independently of whether the country has an initial per capita income of \$5,000 or \$50,000.  It is very hard to provide adequate microfoundations for this type of specifications, explaining why they are rarely used in empirical growth research.

\section{Bounding the effects of the embargo on Cuban GDP}

In this section, we will attempt to provide an upper-bound estimate of the effect of the US trade embargo on Cuban living standards using a range of estimates of the effect of trade openness on economic growth derived from the literature.

As we have shown in the previous section, BGB rely on one estimate, derived from \textcite{Yanikkaya2003}, which is applied to a 12-year horizon from the initial year of the embargo (1960) using equation (\ref{eq1}). In this section, we aim to situate the Yanikkaya estimate in the broader literature on trade and growth, as well as to understand the way in which BGB's conclusions are altered by using other parameter estimates drawn from the literature.

Table \ref{papers} lists some of the most prominent contributions that estimate specifications linking trade openness ratios to per capita income. It is beyond the scope of this note to provide a full systematic survey of estimated effect sizes in this literature. Our more modest purpose is to highlight the values estimated by some of the most well-known contributions in the field and to illustrate the amplitude of the effect sizes they present.  We limit the discussion to papers that use the share of imports and exports on GDP, or some transformation of it, as an explanatory variable.

\textcite{Yanikkaya2003} belongs to a literature that burgeoned in the 1990s using the Barro-style conditional convergence growth equation discussed in the previous section. As we noted there, the inclusion of a conditional convergence term makes coefficient estimates fundamentally different from those of equations without controls for initial income.

\textcite{sala-i-martin2004} provide a useful summary of the results of this literature using a Bayesian averaging of classical estimates (BACE) approach. This approach consists in estimating every combination of possible explanatory variables and studying the resulting distribution of coefficients. BACE methods allow us to observe a full distribution of coefficient estimates across alternative specifications generated within a common estimating framework, permitting comparison of absolute effect-sizes in a way that meta-analyses focused on measures of statistical association, such as that of \textcite{heimberger2022}, are not able to do. They find a steady-state semi-elasticity of income with respect to trade shares of 1.04, around 2.5 times that estimated by Yanikkaya. While the variable fails to satisfy the authors’ strict criteria for a high posterior inclusion probability, it nonetheless displays a sign certainty probability of 94.9\%.

\textcite{frankel_romer_1999} pioneered the use of predicted trade from gravity equations as an exogenous instrument for trade ratios and found a quantitatively large and robust effect of trade on income.  They also found that IV estimates were generally significantly higher than those derived from OLS estimation.  Their baseline estimate is that a one percentage point increase in the trade ratio leads to a 1.97 percent increase in income - nearly five times the semi-elasticity implied by the steady-state level of income derived from the \textcite{Yanikkaya2003} estimates. 

\textcite{alcala_ciccone_2004} focus on the effect of openness on productivity using purchasing-power parity adjusted measures of openness.  Using the same geography-based instruments for trade suggested by \textcite{frankel_romer_1999} as well as additional instruments for institutional quality, they find a statistically and economically significant effect of openness on growth.  In contrast to Frankel and Romer, they use a log-log specification in which log output per worker is regressed on instrumented log of real openness. They estimate an elasticity of output per worker to real openness of 1.23.\footnote{\textcite[629]{alcala_ciccone_2004} also find that, within comparable specifications, the point estimates for the effect of real and nominal openness on growth are very similar.}  This is nearly six times the effect found in \textcite{raghutla2020}.

The above-cited papers use cross-sectional data because the lack of time-variation in geography preclude the use of specifications with country fixed effects.  \textcite{feyrer_2019} addresses this problem by generating a time-varying geographic instrument.  To do so, he relies on the effect of improvements in aircraft technology, which have benefited country pairs with relatively short air routes compared to sea routes. Feyrer uses the resulting heterogeneity to generate a geography-based instrument for trade that varies over time. He estimates an elasticity of income to trade shares of 1.26, very similar to that of \textcite{alcala_ciccone_2004}.\footnote{\textcite{feyrer_2019} derives an estimate of $\beta=0.558$ in the equation $\Delta \ln y_i=\beta \Delta \ln trade_i+\gamma$, which implies an elasticity of income on the trade share of $\frac{\beta}{1-\beta}=1.26$.}

\begin{table}[htpb]
\centering
\small
\resizebox{\textwidth}{!}{
\begin{tabular}{M{30mm}M{15mm}M{22mm}M{18mm}M{25mm}M{15mm}M{15mm}}
\toprule
Authors                                  & Year of Publication & Journal                                              & Years in Sample & Countries Covered                                & Cited by BGB & Citations \\
\midrule
Frankel and Romer                        & 1999                & American Economic Review                     & 1985                                & All countries                                    & No           & 9692      \\
Yanikkaya                                & 2003                & Journal of Development Economics             & 1970-1997                           & Excludes socialist and oil exporting countries   & Yes          & 2025      \\
Sala-i-Martin, Doppelhoffer   and Miller & 2004                & American Economic Review                     & 1960-1996                           & All countries                                    & No           & 2988      \\
Alcala and Ciccone                       & 2004                & Quarterly Journal of Economics               & 1985                                & All countries                                    & No           & 1568      \\
Feyrer                                   & 2019                & American Economic Journal: Applied Economics & 1950-1997                           & Excludes landlocked and oil exporting countries & No           & 660       \\
Raghutla                                 & 2020                & Journal of Public Affairs                    & 1993-2016                           & Brazil, Russia, India, China, and South Africa   & Yes          & 253    \\  
\bottomrule
\end{tabular}
}
\caption{Selected contributions to literature on trade ratios and economic growth}
\label{papers}
\begin{minipage}{\linewidth}
\footnotesize
Citations counts from Google Scholar as of March 29, 2026. 
\end{minipage}
\end{table}

Table \ref{papers} provides some of the basic characteristics of these papers, alongside those of the two contributions—\textcite{Yanikkaya2003} and \textcite{raghutla2020}—cited by BGB. We also provide citation counts and journals of publication as relative proxies for the impact of these papers on the profession.

Of the six papers in the table, the two used by BGB rank third and sixth in terms of total citations. \textcite{frankel_romer_1999} and \textcite{sala-i-martin2004} top the citations counts, while together with \textcite{alcala_ciccone_2004} they are the only papers that appear in general interest economics journals. \textcite{Yanikkaya2003} and \textcite{feyrer_2019} appear in top field journals, while \textcite{raghutla2020} appears in a second-tier journal.  Also worth noting is that Raghutla’s paper uses only data from five BRICS economies, in contrast to all the other papers, which attempt to use the largest samples possible, while in some cases excluding idiosyncratic countries such as socialist or oil-exporting economies.

The purpose of this table is not to argue that any of these papers are superior to the others based on journal reputation or citations. Rather, it is to use generally accepted criteria for measuring impact and standing of intellectual contributions in the profession to identify some of the most prominent contributions in the field. The four papers that we have added to this comparison are well-known, highly cited, and widely seen as having a claim to being seminal contributions to the study of the relationship between development and international trade. Our point is that estimates of the income effects of trade that systematically exclude these contributions cannot be seen as having provided an accurate bound on effect sizes.

\begin{table}[htbp]
\centering
\small
\resizebox{\textwidth}{!}{
\renewcommand{\arraystretch}{1.4}
\begin{tabular}{
>{\raggedright\arraybackslash}p{2.0cm}
M{15mm} M{13mm} M{25mm} M{15mm} M{13mm} M{12mm}M{12mm}M{12mm} M{12mm}M{12mm}M{12mm}
}
\toprule
\multirow{3}{*}{Authors} & \multirow{3}{*}{\makecell[c]{Long-run\\elasticity\\of income\\to trade}}  & \multirow{3}{*}{\makecell[c]{Specification}} & \multirow{3}{*}{\makecell[c]{Dependent\\Variable}} & \multirow{3}{*}{Horizon}
& \multicolumn{3}{c}{\makecell[c]{Growth effect of embargo\\as share of baseline GDP}}
&
\multicolumn{3}{c}{\makecell[c]{Growth effect of embargo \\as share of underperformance \\relative to synthetic control\\(share of log points)\\}} \\
\cmidrule(lr){6-8} \cmidrule(lr){9-11}
& & & & & C1 & C2 & C3 & C1 & C2 & C3 \\
&&&&& (1) & (2) & (3) & (4) & (5) & (6) \\
\midrule
\multirow{2}{*}{Yannikkaya} & \multirow{2}{*}{0.41}
& \multirow{2}{*}{Log-linear}
& \multirow{2}{*}{Growth}
& 12 Years
& 3.8\%
& 8.1\%
& 9.9\%
& 3.1\%
& 6.5\%
& 8.0\% \\

&
& 
& 
& Long-run
& 7.4\%
& 15.9\%
& 19.7\%
& 6.6\%
& 13.6\%
& 16.5\% \\

Raghutla
& 0.19
& Log-log
& Income
& \mbox{Long-run}
& 7.3\%
& 21.9\%
& 34.2\%
& 6.5\%
& 18.2\%
& 27.1\% \\

\makecell[l]{Sala-i-Martin, \\Doppelhoffer\\and Miller}
& 1.04
& Log-linear
& Growth
& \mbox{Long-run}
& 19.7\%
& 45.3\%
& 57.3\%
& 16.6\%
& 34.4\%
& 41.7\% \\

Frankel and Romer
& 1.97
& Log-linear
& Income
& \mbox{Long-run}
& 40.8\%
& 103.2\%
& 136.5\%
& 31.5\%
& 65.3\%
& 79.2\% \\

Alcala and Ciccone
& 1.23
& Log-log
& Income
& \mbox{Long-run}
& 59.4\%
& 269.2\%
& 598.5\%
& 42.9\%
& 120.2\%
& 178.9\% \\

Feyrer
& 1.26
& Log-log
& Income
& \mbox{Long-run}
& 61.5\%
& 282.6\%
& 636.5\%
& 44.1\%
& 123.5\%
& 183.8\% \\
\bottomrule
\end{tabular}
}
\caption{Estimated effects of the embargo under alternative trade-income elasticities}
\label{tab:elasticity_comparisons}
\begin{minipage}{\linewidth}
\footnotesize
Row 1 shows values reported or implied by BGB (see Table \ref{tab:replication}), which use the short-term elasticity $\varepsilon=0.018$ over a 12-year horizon as in equation (\ref{eq1}) and the geometric decomposition of effects relative to the synthetic control given by equation (\ref{thetabgb}). All other rows use the long-run elasticity of income to trade and the logarithmic decomposition of effects relative to the synthetic control given by equation (\ref{share}).
By long-run elasticity of income to trade column we refer to semi-elasticities for log-linear models and use steady-state level of income for growth models with convergence terms.  C1: Trade shock attributed to the embargo equals the 1972 difference between synthetic and historical Cuba trade with rest of world. C2: Trade shock attributed to the embargo equals trade with the US as share of GDP in 1958. C3: Trade shock attributed to the embargo equals C2 plus change in exports between 1958 historical and 1972 synthetic Cuban exports to the United States.  Growth effect of embargo as share of baseline GDP estimated using equations (\ref{eq1})-(\ref{ss}) as appropriate. Growth effect of embargo as share of underperformance divides log change associated with trade shock by log difference of synthetic to historical GDP in 2024. 
\end{minipage}
\end{table}
We now proceed to present estimates of the effects of the US embargo on GDP associated with the elasticities from each of these papers. Our results are presented in Table 2. Following BGB, we present estimates for three scenarios capturing the potential effect of the embargo on trade shares.  We borrow BGB's labels by referring to these scenarios as C1-C3.  Scenario C1 uses the difference between Cuba’s share of trade with the rest of the world in the historical and counterfactual scenarios as the trade effect of the embargo, while scenario C2 uses the decline in trade with the United States from its pre-Revolution level  and C3 adds to this the increase in exports from Cuba to the US that would have been observed in the counterfactual scenario.\footnote{BGB claim that scenario C3 is inconsistent with national accounting identities. We disagree. Scenario C3 simply embodies a decline in the trade share from 55\% to 11\% of GDP. There is no national accounting identity that is inconsistent with Cuba’s trade share falling to 11\% of GDP.}

For each of these scenarios and elasticity estimates derived from the papers in Table \ref{papers}, we present two calculations. The first measures the growth effect of the embargo as a share of baseline GDP, while the second quantifies this growth effect as a share of total underperformance relative to the synthetic control.  

In the first row, we repeat the estimates presented by BGB using the 12-year horizon application of Equation (\ref{eq1}) as well as their calculation of the share of the effect as a share of the synthetic control using the geometric decomposition of (\ref{thetabgb}) — which, as we argued in section \ref{secdec}), is biased down by the omission of the interaction term. This calculation yields growth effects that range from 3.8\% to 9.9\% of baseline GDP (columns 1-3) and from 3.1\% to 8.0\% of the country’s underperformance relative to a synthetic control in 1972 (columns 4-6).

In the second row, we again use the \textcite{Yanikkaya2003} parameters, but instead of using the 12-year horizon adopted by BGB, we use the effects of the assumed shock to openness on the steady-state level of income, as calculated in equation (\ref{ss}). As we can see in columns (1)-(3), the growth effects obtained using steady-state levels of income are approximately twice as large as those calculated by BGB.  

For the purposes of calculating the effects as a share of underperformance shown in columns (4)-(6), we divide the steady-state income effects by the difference between realized and synthetic income in 2024, the latest date for which we can calculate this difference.\footnote{Given that we are using the effect of trade shares on steady-state levels of income, it would make little sense to use underperformance relative to a synthetic control in 1972, as BGB do, as the denominator. We reconstruct the synthetic control estimates up to 2024 using the coefficients reported by BGB in Table 1. For the Cuba historical value, we update their calculation of data-quality-adjusted per capita GDP using the estimates of \textcite{devereux2021}, updated with growth rates of constant-price GDP as reported in the World Bank's World Development Indicators Database.  We note that this adjustment makes our estimates inherently more conservative relative to what would emerge from using 1972 underperformance as a baseline. Note also that for the purposes of the 2024 estimate, the issue of valuing subsidies from the defunct Soviet Union has become immaterial.}  To implement this calculation, we use the additively separable logarithmic decomposition of equation (\ref{share}).\footnote{The effects of just changing the elasticity while keeping BGB's non-separable decomposition are shown in Appendix Table \ref{taba3}.} Moving to the steady-state values while maintaining the use of the Yanikkaya parameters and comparing with Cuba's underperformance as of 2024 also roughly doubles the range of estimates, with the embargo now explaining 7-17\% of Cuban underperformance.

Using the \textcite{raghutla2020} estimates has a similar effect on the share of the country’s underperformance attributable to the embargo for the conservative C1 scenario, but a considerably larger effect for the intermediate and more extreme scenario. The basic reason for this is that the log-log specification implies a proportionately larger effect of larger declines in trade shares, so that while the lower-bound effect is similar to that derived from the \textcite{Yanikkaya2003} parameter estimates, the upper bound, at 27.1\% of Cuba's underperformance, is considerably higher.\footnote{At Cuba's pre-revolution openness of 0.55, Yanikkaya's implied elasticity of steady-state income to openness is  $.23$ which does turn out to be similar to Raghutla's estimated long-term elasticity of $.19$.  Yet this similarity of long-term elasticities does not rescue BGB's claim that the models provide similar predictions, as shown in Table \ref{tab:elasticity_comparisons}. This is because the logarithmic functional form of Raghutla implies a much greater effect of large trade shocks such as those modelled in scenarios C2 and C3.}

These ranges become considerably larger once we turn to using the parameter values estimated in the contributions not cited by BGB. Using the \textcite{sala-i-martin2004} estimates, the growth effect rises to between 17\% and 42\% of Cuba’s underperformance relative to its synthetic control. Using the \textcite{frankel_romer_1999} level specification, in turn, raises the fraction of underperformance explained by the embargo to a range of between 32\% and 79\%. Yet perhaps most remarkable are the results associated with the \textcite{alcala_ciccone_2004} and \textcite{feyrer_2019} parameter estimates, in which the embargo explains between 43\% and 184\% of Cuba’s underperformance relative to a synthetic control.

It is worth pausing for a minute on these last estimates, as attributing to the embargo a share of nearly twice the country’s underperformance may appear excessive. Recall, however, that Scenario C3 was purposely constructed by BGB as an extreme scenario. In our view, a much more reasonable scenario is provided by C2, in which we assume that the embargo reduces the trade share by an amount equal to Cuba’s 1958 trade with the United States.\footnote{BGB argue for treating scenario C1, which uses 1972 values of synthetic and counterfactual world trade, as the baseline, and consider C2 as extreme.  We disagree, for several reasons.  First, the inclusion of Soviet subsidies in measured trade in scenario C1 is inconsistent with their argument that these should be excluded from realized GDP. Second, Cuba has not received Soviet subsidies for nearly four decades, yet it continues to be unable to trade freely with the United States.  To the extent that our aim is to provide estimates that are informative about the causes of Cuba's current economic backwardness, a scenario that values the trade losses to Cuba from the embargo as the foregone trade with the US without adjusting for long-defunct Soviet subsidies is a reasonable baseline.} In this case, the \textcite{alcala_ciccone_2004} and \textcite{feyrer_2019} estimates suggest that, in the absence of the embargo, Cuba’s per capita income would be moderately higher than that of the synthetic control group. For example, the \textcite{feyrer_2019} estimate for scenario C2 would place Cuba’s current GDP per capita at 1.29 times that of the synthetic group, or approximately \$14,800, making it the $11^{\text{th}}$ highest-income economy out of 26 countries in Latin America, with a per capita income between those of Brazil (\$15,507) and Colombia (\$14,483).

\section{Conclusions}

How much of Cuba’s economic underperformance can be traced to the US embargo, and how much to the country’s adoption of centrally planned policies is a highly contentious and disputed issue. BGB attempt to provide an answer using an intuitive approach that combines estimates of the decline in trade generated by the embargo with existing estimates of the elasticity of income with respect to trade openness derived from the empirical literature. Regrettably, their attempt falls short because it relies on parameter estimates that are neither representative nor provide us with reasonable upper bounds for those found in the literature.

Using a range of parameter values that is more representative of key contributions in the literature and adopting a decomposition that does not assume that all interaction effects are attributed to policies, we find that the embargo can explain a large fraction—and in some cases all—of the shortfall in Cuba’s living standards relative to BGB’s synthetic control group. 

As shown in Table 2, this exercise delivers a wide range of estimates, from 7\% to nearly twice the difference between Cuba’s current per capita income and that of the synthetic control group. Even if we exclude the more extreme scenarios and center just on BGB's intermediate scenario, we still get a range of effects of 14\% to 124\% of Cuba's underperformance.  The amplitude of this interval suggests that the range of existing elasticities in the literature makes it difficult to use the 1958 trade shock to establish precise bounds on the embargo's effects.

One possible direction for further research would be to conduct a more systematic survey of estimated elasticities and to focus on central tendency measures or reasonable confidence intervals derived from the distribution of observed estimates. Table \ref{papers} is meant to provide a preliminary illustration of how such an approach could work.

One important area of caution in interpreting these results arises from the fact that estimates of the elasticity of income to trade largely reflect very different types of trade-reducing interventions than those applied to a sanctioned economy. \textcite[311]{rodriguez2000} argue that geographically determined trade may not be a good proxy for trade restrictions imposed by policymakers, because the latter are intentionally designed to achieve specific objectives that may include addressing market failures and fostering growth. Similarly, the effect of geographically-induced trade on income may underestimate the negative effects that sanctions and embargoes have on living conditions when these restrictions are purposely designed to harm the target economy.

An additional complex issue in determining the effects of the US embargo on Cuban living standards relates to the extraterritoriality of US sanctions and their effect in discouraging third countries from trading with Cuba. The recent history of the US's maximum pressure sanctions campaign serves to illustrate the relevance of these effects. In January 2026, President Donald Trump imposed a secondary tariff on any country selling oil to Cuba \parencite{trump2026}. During his first term in office, he also allowed Title III of the Helms–Burton Act—which permits lawsuits in US courts against non-US companies doing business with the Cuban government—to take effect \parencite{segall2019}. There is abundant anecdotal evidence that these measures have deterred key non-US trading partners from engaging with Cuba, suggesting that estimates of sanctions effects that capture only the decline in trade with the United States underestimate the full force of current restrictions \parencite{gordon_2016}.

One possible response to this objection is to argue that BGB's analysis ends in 1989, before the US shift to extraterritoriality in the 1990s with the approval of the Torricelli and Helms-Burton acts.  To the extent that BGB's purpose is only to explain why Cuba was poorer than other countries in the region in the 1980s, that is a legitimate defense.  However, BGB explicitly frame their paper as providing an answer to the question of why Cuba is poor today.  For example, in their introductory paragraph they write: ``Today, Cuba remains a relatively poor nation
due to its slow economic growth after 1959...This
makes Cuba a particularly compelling example of a reversal of fortune ... Understanding its causes is not only interesting for
assessing Cuba’s trajectory but also valuable for economists interested in development
more broadly" [BGB, p. 2]. Regardless of the motivation for BGB's contribution, our interest lies in understanding why Cuba today is a relatively poor country, and the purpose of this comment is to assess the extent to which BGB contribute to providing an answer to this question.

In their Appendix C, BGB present an additional exercise in which they compare Cuba with a synthetic control group of former Communist economies which lost access to Soviet trade subsidies at the same time.  They find that Cuba underperforms the synthetic control group, although the differences in growth trajectories are not statistically significant. While they draw the conclusion that the slower speed of economic reforms in Cuba accounts for the country’s  relative underperformance, we propose an alternative interpretation. In contrast to Cuba, other Soviet Bloc countries did not face a near-total trade embargo from a large, natural trading partner when the Soviet subsidies were withdrawn. It is not hard to imagine the magnitude of the economic collapse that could have ensued in countries like Poland, Hungary, or the Czech Republic if, at the time of the withdrawal of Soviet subsidies, Western Europe had decided to ban trade with them.

While this paper has been critical of BGB’s attempt to bound the effects of the embargo on Cuban GDP, it is worth underscoring that there is much more to their contribution than this exercise. BGB's use of synthetic control methods to chart the potential evolution of the Cuban economy in the absence of sanctions, the embargo, and Soviet subsidies, and their attempt to provide a more realistic estimate of Cuban living standards, are valuable and insightful. BGB provide us with convincing evidence that after 1959 Cuba’s levels of per capita income diverged significantly from those of comparable economies. What they have not done is to satisfactorily disentangle the role that US economic sanctions and domestic economic policies played in accounting for that path.
\newpage
\appendix
\renewcommand{\thetable}{A\arabic{table}}
\setcounter{table}{0}
\section{Appendix}
\subsection{Replication of BGB growth effects}

BGB refer to the existence of a replication code package in their paper. In personal communication, the authors have clarified that such code is not currently being made available to other researchers, pending a decision on publication of the paper. For that reason, we replicate BGB’s calculations based on our best understanding of the explanations provided in the text.

To do so, we use the WebPlotDigitizer app to extract the quantitative values associated with Cuba’s historical and synthetic trade, exports with the United States, and trade with the world from Figures 8, 10, and 12 of their paper. These provide us with a difference between Cuba and synthetic trade with the world of \$530 million in 1972, an initial 1957 value of trade with the United States of \$1,122 million, and a value of trade with the United States in the synthetic control in 1972 which exceeds 1958 exports by \$244 million (all figures measured in 1957 USD).

\begin{table}[htbp]

\centering
\begin{threeparttable}

\caption{Replication of BGB growth effects}

\begin{tabular}{lccc}

\toprule

Scenario & Change in Trade Ratio & Replicated growth effect & BGB reported growth effect \\
\midrule
C1 & 17.1\% & 3.8\%  & 3.8\% \\
C2 & 36.1\% & 8.1\%  & 8.1\% \\
C3 & 44.0\% & 9.9\% & 9.9\% \\

\bottomrule
\label{tab:replication}
\end{tabular}

\end{threeparttable}
\end{table}

Dividing these three figures by the 1958 GDP of \$3.105 billion (in 1957 prices, calculated using \textcite{bolt2023} and US nominal GDP data from \textcite{imf2026}) yields trade ratio variations for Scenarios C1, C2, and C3 of 17.1\%, 36.1\%, and 44.0\%,  respectively. Applying equation (\ref{eq1}) to these values  and using  $\varepsilon = 0.018$, $T_1=1972$, and $T_0=1960$ yields growth effects of 3.8\%, 8.1\%, and  9.9\%, respectively, which are identical to those reported by BGB (See Table \ref{tab:elasticity_comparisons}).

\subsection{Derivation of equation (\ref{thetabgb})}

Exponentiating (\ref{deflny}) and substituting in (\ref{thetaBGB1}):
\begin{equation}
\theta_{BGB}=\frac{e^{\alpha_0+\alpha_1\lambda^{NE}+\alpha_2\eta^S}-e^{\alpha_0+\alpha_1\lambda^{E}+\alpha_2\eta^S}}{e^{\alpha_0+\alpha_1\lambda^{NE}+\alpha_2\eta^{NS}}-e^{\alpha_0+\alpha_1\lambda^{E}+\alpha_2\eta^S}}
    =\frac{e^{\alpha_1(\lambda^{NE}-\lambda^E)}-1}{e^{\alpha_1(\lambda^{NE}-\lambda^E)+\alpha_2(\eta^{NS}-\eta^S)}-1} 
\end{equation}
\begin{equation}
    =\frac{e^{\alpha_1(\lambda^{NE}-\lambda^E)}-1}{(e^{\alpha_1(\lambda^{NE}-\lambda^E)}-1)+(e^{\alpha_2(\eta^{NS}-\eta^S)}-1)+(e^{\alpha_1(\lambda^{NE}-\lambda^E)}-1)(e^{\alpha_2(\eta^{NS}-\eta^S)}-1)}
\end{equation}

 which, using the definitions in (\ref{defs}), gives us (\ref{thetabgb}).

\subsection{Additional Results}
Table \ref{taba3} shows the estimated effects of the embargo calculated for the same elasticity values shown in Table \ref{tab:elasticity_comparisons} but using the non-additively separable geometric decomposition of equation (\ref{thetabgb}) used by BGB.

\begin{table}[htbp]
\centering
\small
\resizebox{\textwidth}{!}{
\renewcommand{\arraystretch}{1.4}
\begin{tabular}{
>{\raggedright\arraybackslash}p{2.0cm}
M{15mm} M{13mm} M{25mm} M{15mm} M{13mm} M{12mm} M{12mm} M{12mm}
}
\toprule
\multirow{3}{*}{Authors} & \multirow{3}{*}{\makecell[c]{Long-run\\elasticity\\of income\\to trade}}  & \multirow{3}{*}{\makecell[c]{Specification}} & \multirow{3}{*}{\makecell[c]{Dependent\\Variable}} & \multirow{3}{*}{Horizon} &
\multicolumn{3}{c}{\makecell[c]{Growth effect of embargo \\as share of underperformance \\relative to synthetic control}} \\
\cmidrule(lr){6-8}\\ 
& & & & & C1 & C2 & C3 \\
&&&&&(1) &(2)&(3)\\
\midrule
\multirow{2}{*}{Yannikkaya} & \multirow{2}{*}{0.41}
& \multirow{2}{*}{Log-linear}
& \multirow{2}{*}{Growth}
& 12 Years
& 3.1\%
& 6.5\%
& 8.0\% \\

&
& 
& 
& Long-run
& 3.8\%
& 8.1\%
& 10.0\% \\

Raghutla
& 0.19
& Log-log
& Income
& \mbox{Long-run}
& 3.7\%
& 11.1\%
& 17.4\% \\

\makecell[l]{Sala-i-Martin, \\Doppelhoffer\\and Miller}
& 1.04
& Log-linear
& Growth
& \mbox{Long-run}
& 10.1\%
& 23.1\%
& 29.2\% \\

Frankel and Romer
& 1.97
& Log-linear
& Income
& \mbox{Long-run}
& 20.8\%
& 52.6\%
& 69.5\% \\

Alcala and Ciccone
& 1.23
& Log-log
& Income
& \mbox{Long-run}
& 30.3\%
& 137.1\%
& 304.8\% \\

Feyrer
& 1.26
& Log-log
& Income
& \mbox{Long-run}
& 31.3\%
& 143.9\%
& 324.1\% \\
\bottomrule
\end{tabular}
}
\caption{Estimated effects of the embargo under alternative trade-income elasticities, geometric decomposition}
\label{taba3}
\begin{minipage}{\linewidth}
\footnotesize
Row 1 shows values reported or implied by BGB (see Table \ref{tab:replication}), which use the short-term elasticity $\varepsilon=0.018$ over a 12-year horizon as in equation (\ref{eq1}) and the geometric decomposition of effects relative to the synthetic control given by equation (\ref{thetabgb}). All other rows use the long-run elasticity of income to trade but maintain  the geometric  decomposition of effects relative to the synthetic control given by equation (\ref{thetabgb}).
By long-run elasticity of income to trade column we refer to semi-elasticities for log-linear models and use steady-state level of income for growth models with convergence terms.  C1: Trade shock attributed to the embargo equals the 1972 difference between synthetic and historical Cuba trade with rest of world. C2: Trade shock attributed to the embargo equals trade with the US as share of GDP in 1958. C3: Trade shock attributed to the embargo equals C2 plus change in exports between 1958 historical and 1972 sythetic Cuban exports to the United States.  Growth effect of embargo as share of baseline GDP estimated using equations (\ref{eq1})-(\ref{ss}) as appropriate. Growth effect of embargo as share of underperformance divides growth effect  as share of initial GDP by ratio of synthetic to historical GDP in 2024. 
\end{minipage}
\end{table}
\newpage
\printbibliography

\end{document}